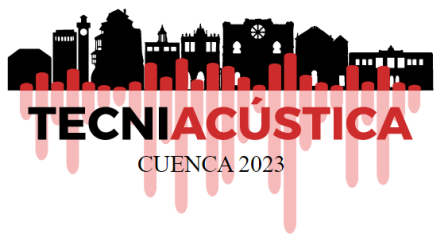

# FOLEY-VAE: GENERACIÓN DE EFECTOS DE AUDIO PARA CINE CON INTELIGENCIA ARTIFICIAL

*Mateo Cámara Largo*[1][*]
*José Luis Blanco Murillo*[1]

[1]Grupo de Aplicaciones del Procesado de Señal. Information Processing and Telecommunications Center. Avda. Complutense 30, ETSI de Telecomunicación, Universidad Politécnica de Madrid

**RESUMEN**

En esta investigación, presentamos una interfaz basada en Autocodificadores Variacionales entrenados con una amplia gama de sonidos naturales para la creación innovadora de efectos de Foley. El modelo tiene la capacidad de operar en tiempo real para transferir nuevas características sonoras a audios pregrabados o voz capturada por micrófono. Además, permite la modificación interactiva de las variables latentes, lo que facilita la realización de ajustes artísticos precisos y personalizados.

Tomando como punto de partida nuestro estudio previo sobre Autocodificadores Variacionales presentado en este mismo congreso el año pasado, profundizamos sobre una implementación existente: RAVE [1]. Nuestro modelo se ha entrenado específicamente para la producción de efectos de audio. Hemos logrado generar con éxito una variedad de efectos de audio que abarcan desde sonidos electromagnéticos, de ciencia ficción, de agua… entre otros muchos, que se publican junto a este trabajo.

Este enfoque innovador ha sido la base de la creación artística del primer cortometraje español con efectos de sonido asistidos por inteligencia artificial. Este hito ilustra de manera palpable el potencial transformador de esta tecnología en la industria cinematográfica, abriendo la puerta a nuevas posibilidades de creación de sonido y a la mejora de la calidad artística en las producciones fílmicas.

**ABSTRACT**

In this research, we present an interface based on Variational Autoencoders trained with a wide range of natural sounds for the innovative creation of Foley effects. The model can transfer new sound features to prerecorded audio or microphone-captured speech in real time. In addition, it allows interactive modification of latent variables, facilitating precise and customized artistic adjustments.

Taking as a starting point our previous study on Variational Autoencoders presented at this same congress last year, we analyzed an existing implementation: RAVE [1]. This model has been specifically trained for audio effects production. Various audio effects have been successfully generated, ranging from electromagnetic, science fiction, and water sounds, among others published with this work.

This innovative approach has been the basis for the artistic creation of the first Spanish short film with sound effects assisted by artificial intelligence. This milestone illustrates palpably the transformative potential of this technology in the film industry, opening the door to new possibilities for sound creation and the improvement of artistic quality in film productions.

***Palabras Clave***— Autocodificador variacional, efectos de audio, aplicación, artes escénicas, inteligencia artificial

## 1. INTRODUCCIÓN

Los efectos de Foley, también conocidos como efectos de sala, se refieren a sonidos que transmiten efectos acústicos cotidianos y que son generados por acciones humanas, ya sea con la ayuda de materiales o sin ellos. Su objetivo es enriquecer la producción audiovisual para hacerla más completa.

Los efectos de Foley son esenciales para construir un ambiente sonoro auténtico y envolvente. Históricamente, las técnicas para implementar efectos de Foley en formato digital han variado según el tipo de sonido que se busca crear. Por ejemplo, para replicar sonidos naturales, es común emplear la síntesis física, que busca simular la fuente del sonido mediante ecuaciones que describen su comportamiento físico [2], o incluso, reproducir a pequeña escala el ambiente o los componentes físicos implicados.

Para el audio espacial, por otro lado, se utiliza la síntesis modal, donde una señal de entrada con características ruidosas se procesa para imitar modos de resonancia específicos [3]. Además, a lo largo de los años, se han desarrollado técnicas, como la síntesis aditiva, sustractiva, por muestreo... etc. Sin embargo, son técnicas que presentan limitaciones a la hora de generar sonidos complejos. O bien requieren una gran cantidad de osciladores en paralelo que incrementan el coste computacional exponencialmente, o bien son incapaces de capturar la riqueza tímbrica por falta de variedad y calidad de las muestras disponibles.

Conocidas estas limitaciones surge una pregunta: ¿cómo proceder cuando se quiere inventar un sonido completamente nuevo, que no tenga un referente físico claro, que tenga calidad y que, además, sea fácil e intuitivo de manipular? Aquí es donde entra en juego la síntesis generativa apoyada por modelos de inteligencia artificial. Esta aproximación ofrece una solución innovadora, cuyos módulos pueden aprender de vastos conjuntos de datos, reconocer patrones y características clave de diferentes sonidos y, finalmente, generar audio que es tanto innovador en su contenido como coherente con las expectativas del oyente.

En este estudio presentamos el Foley-VAE, una aplicación basada en Autocodificadores Variacionales (VAE) [3] diseñada específicamente para apoyar la generación innovadora de efectos de Foley. Describimos el proceso tanto artístico como técnico que hemos llevado a cabo, desde la identificación de la necesidad de un efecto particular, pasando por el preprocesamiento de una base de datos con sonidos de referencia y el entrenamiento con un sistema generativo (RAVE [1] en nuestro caso), hasta el posprocesado del sonido generado. El esquema desarrollado incluye una interfaz en MAX 8 que cualquier usuario puede descargar para interactuar y disfrutar de nuestro sistema. Esta aplicación se ha utilizado para apoyar una creación artística que se espera presentar al público en los próximos meses.

## 2. OTROS TRABAJOS RELACIONADOS

### 2.1 Síntesis de efectos de Foley

La síntesis de efectos Foley ha evolucionado a lo largo del tiempo. Originalmente se centraba en replicar sonidos a través de modelos físicos que emulaban efectos como arañazos, rodaduras, raspados o golpes [5]. Junto con estos, la síntesis de sonidos naturales, tales como truenos, agua, pasos, fuego y lluvia, han sentado las bases de la Foley digital en décadas recientes [6][7]. Sin embargo, estos últimos todavía plantean importantes retos en cuanto a la fidelidad, lo que explica que se sigan empleando estudios para la recogida de grabaciones.

El interés en los efectos de Foley es evidente a la vista del número de competiciones que premian las mejores soluciones científicas. Entre las competiciones más relevantes y conocidas se incluyen Blizzard Challenge, ChiME, DCASE, Music Demixing Challenge, o el AI Song Contest [8]. Estas competiciones favorecen el desarrollo de nuevas técnicas, entre las que están proliferando las basadas en esquemas de aprendizaje profundo.

### 2.2 Aprendizaje profundo y procesado de señal acústica

El aprendizaje profundo ha impactado significativamente en multitud de áreas, siendo el tratamiento de señales acústicas una de las más beneficiadas. Durante los últimos años hemos asistido un aumento continuado en el número de investigaciones [9] que recurrentemente nos presentan nuevos esquemas y nuevas arquitecturas. En este contexto, WaveNet [10] se posiciona como una referencia esencial. Representa uno de los primeros modelos generativos que manejan audio desde su forma de onda optando por un enfoque autorregresivo. En este método, la predicción de una muestra está influenciada por las predicciones previas, lo que resulta en un sonido coherente y fluido. Sobre estos mismos fundamentos han surgido esquemas como AudioLM [11], orientado a la generación de sonidos, y MusicLM [12], enfocado en la producción musical.

Por el contrario, son muchos los trabajos que descansan sobre representaciones espectrales [1][13]. La principal diferencia es que extraen la información acústica del espectrograma de la señal, aprovechando así una representación condensada de la información de amplitud a lo largo del tiempo y de la frecuencia. Este esquema evita utilizar la forma de onda en crudo, lo cual puede ser computacionalmente costoso dado el gran número de muestras necesario para representar la señal de audio en tiempo. Esta estrategia es la que se ha seguido en este trabajo, puesto que es especialmente útil para trabajar sobre los efectos de audio [8][14]. Sin embargo, esta estrategia presenta algunas desventajas, como la pérdida de detalles finos o la dificultad de reconstrucción de la señal a una nueva forma de onda con la calidad sonora deseable.

Algunos antecedentes recientes han demostrado que el aprendizaje no supervisado es capaz de generar representaciones acústicas convincentes a través de autocodificadores [15]. Sin embargo, el espacio latente que resulta del entrenamiento de los esquemas clásicos no posee de forma natural las propiedades necesarias para generar nuevos sonidos. Es decir, carece de verdadera capacidad generativa. La introducción de los esquemas variacionales, como el VAE, infunde propiedades de normalidad al espacio latente, dotando al autocodificador de las deseadas capacidades generativas. Los detalles matemáticos de estos esquemas y sus capacidades se discutieron el año pasado en este congreso, partiendo de un esquema diseñado para profundizar en la estructura de estos esquemas y las propiedades del espacio latente [16].

Aunque también existen otros enfoques, como el Procesado de Señal Digital Diferenciable (DDSP), que ha ganado atención por su capacidad para modelar y replicar efectos de audio con alta precisión [30], los VAEs se destacan especialmente en la generación y manipulación de efectos de audio. Mientras que DDSP es excepcional en la réplica precisa de efectos de audio y ha sido estudiado para simular efectos analógicos [31], los VAEs ofrecen una mayor flexibilidad para la creatividad y la experimentación, estableciendo así un equilibrio único entre fidelidad y exploración en el diseño de efectos de audio.

El avance que han supuesto los esquemas variacionales también ha desencadenado toda una serie de aplicaciones en áreas como la generación musical [17], el modelado analógico virtual [18] y la síntesis de audio [19] entre otros. En épocas recientes, las propuestas más innovadoras para crear efectos de audio vienen empleando técnicas de aprendizaje profundo y modelos generativos como los VAEs [20]. Estos se pueden potenciar con modelos de difusión, que aseguran la coherencia temporal de las muestras. En esta misma línea, trabajos como AutoFoley [14], FoleyAutomatic [2] o Deep-Modal [21] son ejemplos de condicionamiento de la generación de audio utilizando señales de vídeo, simulando el trabajo de un artista de Foley cuando sincroniza los efectos con la imagen. Paralelamente, se han desarrollado tendencias que adaptan modelos concebidos para música al ámbito de los efectos de audio [22], o que adoptan enfoques adversarios para la síntesis [23].

Un ejemplo reseñable es el autocodificador de tiempo real RAVE [1]. Este esquema propone un sistema generalista que no está limitado únicamente a la música (generación de instrumentos, voces, pasos…), y que utiliza representaciones espectrales de audio para el entrenamiento y la generación un espacio latente con las características de los audios provistos. Ha demostrado su eficacia en la generación de efectos de audio y, gracias a su integración con redes adversarias, logra una reconstrucción fiel del audio original [37].

Esta plataforma ya ha sido exhaustivamente evaluada y ha servido de inspiración para múltiples investigaciones [13][27]. Hay que tener en cuenta que, para poderse utilizar como sintetizador, se debe reconstruir la fase de la señal con precisión. La gran mayoría de los sintetizadores analizados utilizan Hi-Fi GAN [28] como modelo de reconstrucción, aunque ya comienza a haber trabajos que introducen este tipo de información dentro del propio espacio latente [29].

## 3. ESTRUCTURA E IMPLEMENTACIÓN

Foley-VAE es una aplicación pensada para la generación de efectos de Foley mediante inteligencia artificial. Puesto que el abanico de efectos de Foley es virtualmente infinito, este sistema se centra en optimizar y enriquecer sonidos específicos preseleccionados. No es una herramienta para concebir efectos completamente nuevos de forma autónoma, sino que potencia la labor artística del artista de Foley, facilitando el proceso de exploración de las posibles variantes. En este capítulo presentamos el flujo de trabajo, que involucra el preprocesamiento de los audios, la adaptación del modelo de inteligencia artificial y, finalmente, la generación del efecto sonoro. El mismo flujo que describimos puede seguirse para desarrollar nuevos efectos o crear nuevas combinaciones.

### 3.1 Preprocesado

Una vez el artista de Foley ha seleccionado un conjunto de sonidos base adecuados para cierta escena, el preprocesado de los datos puede comenzar. En primer lugar, se segmentan los audios de longitud arbitraria en segmentos de 5 segundos y se homogeniza la frecuencia de muestreo a 44,1 kHz, que es la frecuencia de trabajo de RAVE. Posteriormente, a cada uno de estos segmentos se les aplica individualmente una gama de efectos preexistentes con parámetros aleatorizados. Esto significa que cada fragmento sonoro experimenta solo una transformación específica. Con esto generamos un conjunto de variaciones sobre el sonido original que enriquezcan la base de datos original, y que se utiliza como conjunto de entrenamiento de la IA. La lista de efectos, así como las variaciones que hemos seleccionado puede encontrarse en la Tabla 1. Se trata de cinco efectos sonoros básicos ampliamente utilizados en la industria cinematográfica. Estos se computan automáticamente con el software Reaper[2], utilizando un script basado en el lenguaje LUA[3].

El siguiente paso es adaptar estos registros para procesarlos con nuestro VAE. Específicamente, la amplitud de la forma de onda se normaliza para que varíe entre 1 y -1 y se fragmenta el registro en $N$ señales submuestreadas. Este submuestreo es conocido como descomposición multibanda y permite obtener diferentes precisiones en el espectro, favoreciendo nuestra capacidad para representar detalles finos y gruesos al mismo tiempo [33]. Con esto conseguimos

---

[2] https://www.reaper.fm/

[3] https://github.com/MateoCamara/Reaper-automatic-random-generation

una base de datos de gran riqueza para el entrenamiento de una red neuronal.

**Tabla 1**. Lista de efectos y rangos aplicados

| Efecto | Parámetros | Rango |
|---|---|---|
| Chorus | Longitud (ms) | 0 – 200 |
| | Voces | 0 – 8 |
| | Wet Mix (dB) | 30 – 42 |
| | Dry Mix (dB) | 30 – 42 |
| | Ratio (Hz) | 0 – 16 |
| | Cambio de tonos | 0 – 1 |
| Distorsión | Ganancia (dB) | 10 – 40 |
| | Dureza | 1 – 10 |
| Ecualizador | Ganancia 1 (dB) | -5 – 1 |
| | Ganancia 2 (dB) | -5 – 5 |
| | Ganancia 3 (dB) | -5 – 5 |
| | Ancho Banda 1 (oct) | 0,1 – 4 |
| | Ancho Banda 2 (oct) | 0,1 – 4 |
| | Ancho Banda 3 (oct) | 0,1 – 4 |
| Rever | Wet Mix (dB) | 0 – 3 |
| | Dry Mix (dB) | 0 – 3 |
| | Habitación | 30 – 90 |
| | Dampening | 0 – 100 |
| | Paso bajo | 0 – 10mil |
| | Paso alto | 10mil – 20mil |
| Flanger | Longitud (ms) | 0 – 200 |
| | Realimentación (dB) | -120 – 6 |
| | Wet Mix (dB) | -30 – 12 |
| | Dry Mix (dB) | -30 – 12 |
| | Ratio (Hz) | 0 – 100 |

**3.2 Entrenamiento de la Inteligencia artificial**

Como se ha señalado, el sistema de inteligencia artificial que empleamos en este trabajo se basa en el VAE de RAVE [1]. El sistema utiliza dos etapas para su entrenamiento. En primer lugar, se entrena minimizando la distancia de la señal de entrada y la reconstruida sobre las señales multiescala según se propone en [33]. De esta forma, se minimiza la diferencia en la amplitud de las señales y no se penaliza al VAE a la hora de reconstruir la fase, que potencialmente puede no tener interés perceptual y puede representar un reto aun mayor para la correcta reconstrucción. Además de este error, se incorpora la pérdida del codificador y decodificador derivada de la naturaleza gaussiana del VAE, que se minimiza a través del *Evidence Lower Bound* (ELBO) [4].

La segunda fase consiste en un refinamiento de los parámetros usando técnicas adversarias. Los creadores de RAVE sostienen que este paso potencia la naturalidad y calidad del audio generado [1]. En esta etapa se detiene el entrenamiento del codificador y solo se modifican los parámetros del decodificador. Puesto que el espacio latente se ha definido en la etapa anterior, éste puede tomarse como la distribución base para una Red Generativa Adversaria (GAN) [34]. Es decir, se utilizan los audios decodificados procedentes del muestreo del espacio latente y se comparan con los sonidos originales utilizando un discriminador. Este discriminador tratará de discernir si el audio que se le presenta proviene del muestreo del VAE o si se trata de un audio genuino. De esta forma, cuando el esquema adversario sea incapaz de determinar si se trata de uno o de otro significará que el VAE está bien entrenado, hasta el punto de generar audios verdaderamente realistas.

Al término de estas dos etapas, el VAE está listo para generar audios similares a los originales, así como mezclas inteligentes de los anteriores. Como parte del entrenamiento, se añade un procesamiento más para descartar variables latentes que no aporten información relevante. Esto se logra descartando las que se asemejen excesivamente a la prior gaussiana del VAE. Adicionalmente, en RAVE se proyectan las variables que contienen información a un subespacio menor que concentre las características similares en una misma variable y separe las características diferentes en variables incorreladas. Este último paso permite un uso interactivo más sencillo e intuitivo del autocodificador durante la fase de generación.

**3.3 Generación**

Para la generación de audio se ha diseñado una interfaz interactiva. Ésta se ha desarrollado en MAX 8[4], un software visual que permite la creación de interfaces para audio y que soporta el uso de VAEs en tiempo real. La Imagen 1 muestra una vista previa de la interfaz diseñada. Una versión *standalone* se encuentra disponible en abierto a través de la web[5].

La interfaz permite el uso del VAE de dos formas diferentes. Por un lado, permite al usuario cargar un sonido arbitrario y utilizarlo como fuente de excitación del VAE. Es decir, introducir el sonido en el codificador, obtener una muestra en el espacio latente y decodificarlo. El sonido resultante será una modificación del timbre del original con respecto de los audios con los que se entrenó el VAE. Anecdóticamente se permite que la señal de entrada al VAE sea la entrada de micrófono, pudiendo así hacer pruebas con la voz, palmadas o silbidos, con resultados habitualmente extravagantes.

La segunda modalidad permite asignar valores exactos a las variables en el espacio latente. De esta forma se puede hacer una exploración explícita de los audios que puede crear el VAE. Este es el uso recomendado para la generación de efectos de Foley. El usuario puede activar y desactivar las variables del VAE, modificando su valor y comprobando el sonido producido a través de la interfaz interactiva. Cuando ha encontrado un valor razonable, puede descargarlo y utilizarlo en su producción.

---

[4] https://cycling74.com/

[5] https://mateocamara.com/resources/

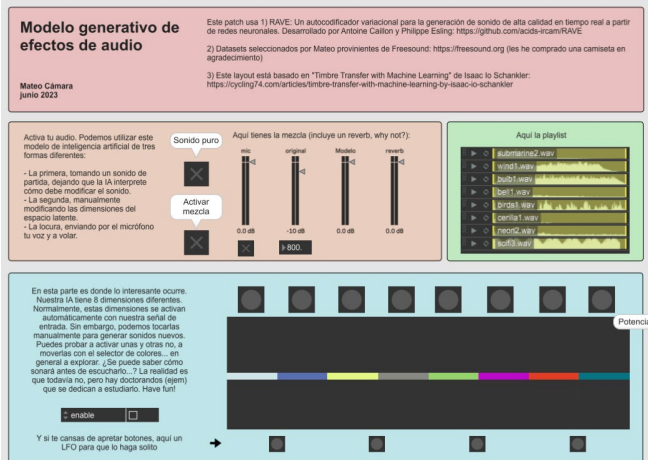

**Imagen 1**. Vista previa de la interfaz de Max 8 en la que se encuentra embebido el VAE.

Adicionalmente, se han incluido módulos de procesado digital más clásicos dentro de la interfaz diseñada, para complementar el flujo de procesado. Estos módulos permiten combinar las capacidades avanzadas del VAE con elementos paramétricos más controlables y conocidos. Se ha incluido un filtro paramétrico, que permite modificar la estructura frecuencial del audio generado por el VAE, modulaciones en amplitud basadas en LFOs y modulaciones en amplitud constantes que permiten simular golpeos o encendidos. Aunque no se trata de un conjunto de herramientas exhaustivo, son las suficientes para demostrar cómo podemos integrar en el esquema propuesto las capacidades avanzadas de la inteligencia artificial y las más clásicas propias de las estaciones digitales de trabajo. Con todo esto, la herramienta implementada permite diseñar eficientemente una gran variedad de efectos de Foley. No obstante, siempre se puede continuar el procesado en nuestro software de edición preferido una vez hemos generado un audio base satisfactorio.

Por último, La interfaz permite agregar VAEs adicionales. Estos componentes adicionales se pueden utilizar para sintetizar a la vez diferentes sonidos y realizar una mezcla en tiempo real. Los audios generados por separado pueden combinarse de forma aditiva, por lo que se da lugar a relaciones entre sonidos o mezclas que físicamente sean imposibles. Por otro lado, la combinación puede realizarse directamente en el proceso de decodificación del VAE, por lo que ambas señales excitarían al mismo tiempo el espacio latente y este se descodificaría como si se tratase de una única señal. De esta forma, se logra un nivel de abstracción superior en el sentido de que se pueden entrenar dos modelos por separado, con tipología de audio muy dispar, y generar una mezcla coherente. Sin embargo y a pesar de su potencial, está técnica aún está siendo refinada, ya que tiende a ser inestable y computacionalmente costosa, por lo que su aplicación práctica sigue siendo un desafío.

## 4. EXPERIENCIAS

En esta sección, exploramos la eficacia de nuestro modelo en la generación de efectos de sonido de pasos. Este tipo de efecto es ideal para una evaluación objetiva, ya que es universalmente conocido y hay abundantes bases de datos disponibles para comparar los resultados. Los efectos de pasos también ofrecen la flexibilidad de simular tanto escenarios realistas como imaginativos, como, por ejemplo, el sonido de un paso en una superficie que sea una mezcla entre metal y agua. Además, los efectos de pasos son fáciles de alinear entre sí, ya que suelen ser explosiones de energía fácilmente identificables en el audio.

### 4.1 Materiales y entrenamiento

Para este estudio hemos utilizado una base de datos de pasos grabados en un estudio de Foley. Hay aproximadamente 15.000 registros de audio grabados con 16 bits de precisión a 48 kHz de frecuencia de muestreo y una duración aproximada de 30 segundos. Cada uno contiene un tipo diferente de zapatos, materiales y cadencia de paso. Para este estudio se han agrupado los sonidos en función del tipo de material en seis categorías: madera, metal, roca, tela, tierra y otros (agua, nieve…). Cada grupo contiene 120 audios y una duración total de 60 minutos.

Hemos entrenado el VAE tomando un 90% de muestras para entrenamiento y un 10% para test y validación. Para el entrenamiento se ha utilizado Pytorch 2.0.1, con CUDA 11.7 en un sistema UBUNTU 22.04 con aceleración por Hardware en GPU modelo NVIDIA GeForce RTX 3090. El modelo de nuestro VAE se entrenó durante 24 horas para evaluar los resultados del proceso de entrenamiento. El modelo final corresponde con el punto de control con menor error de validación, que se produjo a las 3 horas del entrenamiento. La Imagen 2 muestra la evolución del error (detallado en la Sección 3.2) en función del número de muestras del conjunto de entrenamiento computadas.

### 4.2 Espacio latente

La información sonora se encuentra embebida en un espacio latente de 128 variables. El sistema RAVE propone un método de reducción de dimensionalidad mediante la técnica de PCA sobre un subespacio de 8 variables que contienen el 95% de la información acústica (fidelidad) embebida en el espacio original. Sobre este subespacio, la Imagen 3 presenta una reducción a dos dimensiones mediante la técnica t-SNE. Esta reducción permite tener en cuenta las características del espacio latente y ofrecer una visualización interpretable. La imagen muestra el subconjunto de test embebido en el espacio latente en función de los distintos tipos de materiales considerados; madera, roca, tela, tierra, metal y otros.

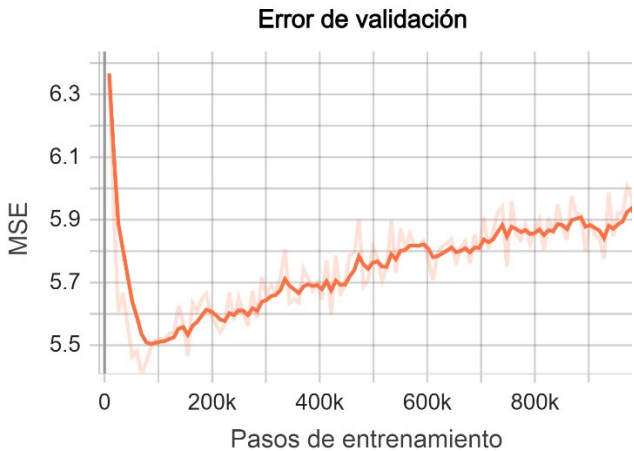

**Imagen 2**. Evolución del error de validación a lo largo del número de actualizaciones de pesos de la red.

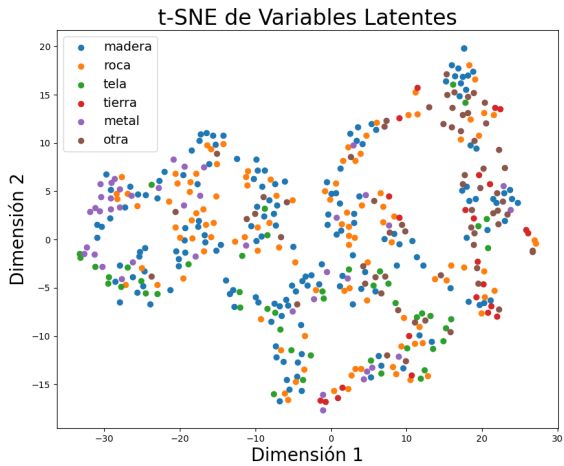

**Imagen 3.** Representación t-SNE del espacio latente del VAE.

En la Imagen 3 se observa un ligero agrupamiento de los diferentes tipos de materiales. La madera (azul) y la roca (naranja) se han agrupado juntos en la región superior de la dimensión 2. El motivo es que se trata de sonidos cercanos, percusivos, secos, de poco transitorio. La tela (verde) ha quedado en la región inferior de la dimensión 2, claramente separada de los otros dos tipos de materiales citados, al tratarse de sonidos suaves, más parecidos a un rozamiento que a un golpeo. El metal (morado) ha quedado claramente separado del resto de materiales en la región izquierda, siendo el sonido con mayor transitorio. Mientras, la tierra (rojo) y resto de materiales (marrón), con sonidos acuosos, de timbre muy diferente al resto, quedan en la región derecha. La principal conclusión es que el VAE ha sido correctamente entrenado, puesto que el espacio latente parece que logra distinguir los diferentes detalles y timbres de los audios estudiados. Sin embargo, la estructura del t-SNE parecería indicar que las clases se entremezclan en el espacio latente. La primera conclusión debe corroborarse evaluando la calidad de los audios generados. La segunda analizando la estructura del espacio latente en función de sus variables.

### 4.3 Capacidad de generación

En esta sección se mide la capacidad de generación desde dos perspectivas. La primera es la capacidad de regeneración de un sonido. Es decir, la capacidad de recrear un sonido lo más parecido al original. Este estudio permite computar el error objetivo cometido para un registro de referencia. La segunda es la capacidad de mezcla de dos sonidos. Es decir, la generación de un audio totalmente nuevo, tomando como referencia dos audios originales. En este caso no se puede hacer una valoración objetiva al carecer de una referencia, por lo que se realiza una valoración de referencia incompleta. En todo caso, se invita al lector a escuchar y evaluar por sí mismo la calidad de los audios adjuntos a este trabajo[6].

#### 4.3.1 Regeneración del audio

Para evaluar la eficacia de la regeneración, se toma cada audio del conjunto de test y se procesa a través del modelo. Esto da como resultado una versión generada para cada audio original. Empleamos tres métricas diferentes para comparar la calidad del audio generado con el original: el Error Cuadrático Medio (ECM), la métrica de calidad objetiva ViSQOL, que realiza una estimación sobre la subjetiva [35], y la distancia de Fréchet (FAD), que evalúa la calidad sin necesidad de referencia [36]. Estas métricas se calculan para todo el conjunto de audios originales. Los resultados de esta evaluación se resumen en la Tabla 2.

**Tabla 2.** Resultados de regeneración de audio en el VAE.

|  | ECM | ViSQOL | FAD |
|---|---|---|---|
| Valor | $51 \pm 27$ | $4,1 \pm 0,2$ | $\sim 0$ |

En esta tabla se pueden observar los resultados agregados para todos los materiales. La métrica subjetiva ViSQOL, que toma valores entre 1 (peor calidad) y 5 (mejor calidad), se encuentra cerca del máximo. La distancia FAD es mínima, indicando que los sonidos producidos son efectivamente pasos y no ruido. Por último, el valor de ECM, calculado sobre el espectrograma MEL, no demuestra la calidad del sonido, pero puede tomarse como referencia para futuros estudios. La imagen 4 muestra la distribución de valores obtenida para en ECM. El error es análogamente bajo en más del 85% de los registros.

---

[6] https://mateocamara.github.io/foley-vae/

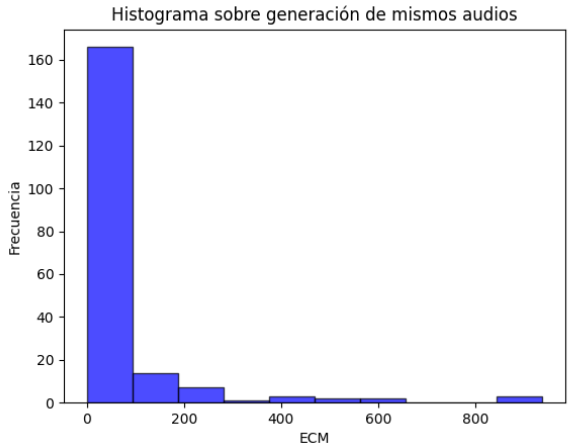

**Imagen 4.** Distribución de errores cuadráticos medios de la regeneración de audio en el VAE.

La Imagen 5 muestra una matriz de las distancias de Fréchet (FAD) clasificadas según el tipo de material. En general, los valores cercanos a cero indican que el audio reconstruido es de buena calidad. Hay tres aspectos que resaltan:

1. La diagonal principal de la matriz, que compara cada audio generado con su versión original, tiene valores muy bajos. Esto significa que el modelo es consistentemente bueno en recrear los sonidos originales.
2. En particular, la columna que corresponde a los registros de tipo "tierra" es especialmente singular. Aunque los valores más altos de FAD en esta columna sugieren que estos sonidos tienen un timbre muy distintivo, el bajo valor de FAD entre el sonido original y el reconstruido de "tierra" indica que el modelo VAE ha sido eficaz en capturar estos detalles únicos.
3. En general, valores bajos en toda la matriz (sin contar con la diagonal principal) indica que los sonidos son relativamente parecidos entre sí. Solamente detalles pequeños los diferencian.

### 4.3.2 Mezcla de audios

Para la generación de mezclas se han tomado dos audios de pasos sobre materiales diferentes. Se han sumado las representaciones latentes de cada uno de ellos y el resultado se ha decodificado. El audio generado es una mezcla inteligente de ambos. En la Imagen 6 se muestra la distancia de Fréchet para las mezclas de audios. El eje de abscisas contiene uno de los materiales usados en la mezcla, y el eje de ordenadas los sonidos sin mezcla originales.

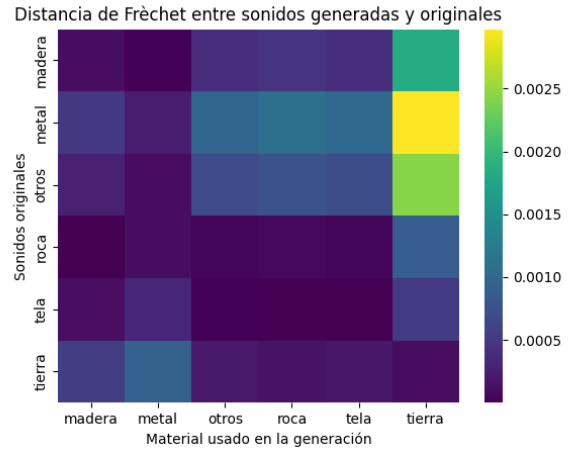

**Imagen 5.** Matriz de valores de FAD en función del tipo de material generado y comparado sobre el conjunto original

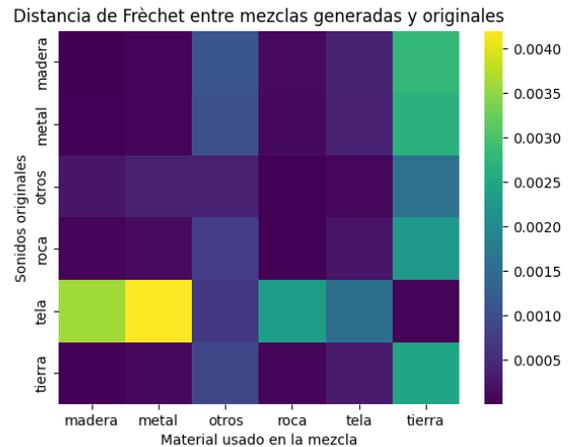

**Imagen 6.** Matriz de valores de FAD sobre audios mezcla en función del tipo de material.

Estas observaciones ofrecen una visión útil sobre cómo diferentes tipos de sonidos interactúan en las mezclas y cómo el modelo es capaz de mantener la coherencia con la base de datos original. En este caso, podemos señalar varios aspectos clave relacionados con los valores de la distancia de Fréchet (FAD) y las mezclas de sonidos:

1. Los valores bajos de FAD indican que las mezclas de sonidos resultan en efectos de pasos que son coherentes con la base de datos original. Esto sugiere una buena calidad en la generación de sonidos.
2. Se nota que ciertos tipos de sonidos son más dominantes que otros en las mezclas. Por ejemplo, cuando se mezcla cualquier sonido con "tela," los detalles tienden a desaparecer y el sonido de la tela queda subyugado por el otro sonido con el que se combina. En contraste, el sonido de "tierra" es muy dominante y tiende a sobresalir en cualquier mezcla en la que participa.

## 5. CONCLUSIONES

En este trabajo se ha presentado un método sistemático para la generación de efectos sonoros mediante el uso de autocodificadores variacionales: Foley-VAE. Se ha analizado en detalle la investigación sobre RAVE y se ha desarrollado una aplicación para facilitar su uso. Los resultados se han utilizado en un contexto real para generar efectos sonoros en uno de los primeros cortometrajes realizados en España en beneficiarse de la inteligencia artificial para la creación de efectos de Foley. Además, se ha presentado y analizado un generador de pasos, que se ha tomado como referencia para analizar la capacidad generadora del sistema. Esta aplicación se ha puesto a disposición del público para su uso libre.

Los resultados presentados sobre los efectos sonoros de pasos han demostrado que Foley-VAE es un sistema que puede generar sonidos de gran calidad. Cuando se trata de regenerar los mismos audios contenidos en el subconjunto de test, la calidad perceptual equivalente es muy elevada. Cuando se trata de evaluar mezclas de sonidos, las nuevas creaciones son estadísticamente similares a los sonidos originales, demostrando capacidad generativa coherente. También resulta muy interesante analizar el espacio de embebimiento y el comportamiento de los distintos materiales.

Se ha comprobado que el esquema VAE entrenado ha adquirido conocimientos sobre la naturaleza de estos, hasta el punto de que las mezclas de materiales que físicamente no son posibles, son perceptualmente coherentes. También ha podido comprobarse la dominancia de unos materiales sobre otros, que intuitivamente cuadra con lo que cabría esperar (por ejemplo, el sonido acuoso de una pisada sobre barro dominando al de una tela). Recomendamos a los lectores que escuchen por sí mismos los resultados para valorar subjetivamente estas conclusiones.

Como línea de trabajo futura se contempla añadir una etapa adicional al proceso de inferencia del VAE que permita desentrelazar las variables latentes, tratando de garantizar la independencia entre sí, para así poder identificar unívocamente las características acústicas de cada variable. Esta mejora permitiría un uso más intuitivo de la aplicación, minimizando modificaciones arbitrarias de las variables hasta dar con un efecto deseado. Por otro lado, se contempla también utilizar modelos de difusión sobre el espacio latente que permitan mantener la coherencia temporal del sonido. Ello facilitaría la labor de introducción del efecto de audio en la producción artística. Por último, cabe analizar las ponderaciones de mezclas para entender la dominancia de unos efectos sobre otros y armonizarlas entre sí, todo ello trabajando sobre el espacio latente y en base a sus propiedades y estructura propias.

## 6. AGRADECIMIENTOS

Los autores agradecen a A. Kampmann, como director del cortometraje "El Testigo", por su confianza en la aplicación de técnicas de inteligencia artificial en su obra cinematográfica, así como a Pancho Aguirre, como técnico de sonido, quien ha sido uno de los primeros usuarios de nuestra herramienta.

Este trabajo ha sido financiado conjuntamente por el Ministerio de Economía y Competitividad del Gobierno de España dentro del proyecto PID2021-128469OB-I00, el Programa de Investigación e Innovación de la Unión Europea Horizon 2020 dentro del "Grant Agreement No. 101003750", y el Programa propio de Investigación de la Universidad Politécnica de Madrid.

## 7. REFERENCIAS